# Fe$_3$O$_4$@astragalus polysaccharide core-shell nanoparticles for iron deficiency anemia therapy and magnetic resonance imaging *in vivo*

*Kai Wang, Lina Li, Xiaoguang Xu\*, Liying Lu, Jian Wang, Shuyan Wang, Yining Wang, Zhengyu Jin, Jin Zhong Zhang, Yong Jiang \**

*K. Wang, Prof. X. G. Xu, Prof. Y. Jiang. Beijing Advanced Innovation Center for Materials Genome Engineering, School of Materials Science and Engineering, University of Science and Technology Beijing, Beijing 100083, China*

*E-mail: xgxu@ustb.edu.cn (X.G. Xu), yjiang@ustb.edu.cn (Y. Jiang)*

*Prof. L. N. Li, S. Y. Wang. School of Chinese Medicine, Beijing University of Chinese Medicine, Beijing 100029, China*

*Prof. L.Y. Lu. School of Chemistry and Biological Engineering, University of Science and Technology Beijing, Beijing 100083, China*

*Dr. J. Wang, Prof. Y. N. Wang, Prof. Z. Y. Jin. Department of Radiology, Peking Union Medical College Hospital, Chinese Academy of Medical Sciences, Beijing 100730, China*

*Prof. J. Z. Zhang. Department of Chemistry & Biochemistry, University of California, Santa Cruz, CA 95064, USA*

ABSTRACT: $Fe_3O_4$@astragalus polysaccharide core-shell nanoparticles ($Fe_3O_4$@APS NPs) were demonstrated to be an efficient therapeutic drug for treating iron deficiency anemia (IDA) *in vivo*. The $Fe_3O_4$@APS NPs have been synthesized using a two steps approach involving hydrothermal synthesis and subsequent esterification. Transmission electron microscopy (TEM) and Fourier transform infrared (FTIR) spectroscopy studies show that APS are attached on the surfaces of the highly monodisperse $Fe_3O_4$ NPs. Dynamic light scatting (DLS) and magnetic characterizations reveal that the $Fe_3O_4$@APS NPs have outstanding water solubility and stability. Cytotoxicity assessment using Hela cells and pathological tests in mice demonstrate their good biocompatibility and low toxicity. The IDA treatment in rats shows that they have efficient therapeutic effect, which is contributed to both the iron element supplement from $Fe_3O_4$ and the APS-stimulated hematopoietic cell generation. Moreover, the $Fe_3O_4$@APS NPs are superparamagnetic and thus able to be used for magnetic resonance imaging (MRI). This study has demonstrated the potential of nanocomposites involving purified natural products from Chinese herb medicine for biomedical applications.

Iron is one of the most important trace elements in human body and is essential for the normal function of organisms, particularly for the metabolism and immune functions.[1] As reported by World Health Organization, 2 billion people have anemia, with nearly 1 billion suffering from iron deficiency anemia (IDA).[2] Therefore, IDA is a common nutritional disease in modern society.[3] Oral iron supplementation, such as ferrous salts, is common in the treatment and prevention of iron deficiency in human body.[4] However, it usually causes side effects, such as epigastric pain, diarrhea and constipation.[5] Therefore, it is highly desired to develop new iron supplements with no or few side effects.[6]

Ferumoxide, known as Feridex in America and Endorem in Europe, is a colloid of ultra-small superparamagnetic iron oxide (USPIO) nanoparticles (NPs) coated by dextran.[7] It is a FDA-cleared drug for the treatment of IDA in adult chronic kidney disease patients.[8] [9] After intravenous administration, the USPIO NPs are cleared from blood by phagocytosis.[10] The USPIO NPs are metabolized in the lysosomes into a soluble, nonmagnetic form of iron that becomes part of the normal iron pool after the intracellular uptake. Among the USPIO NPs, $Fe_3O_4$ NPs have been proved to exhibit good superparamagnetic property below 15 nm and can be used as both an iron supplement and a magnetic resonance imaging (MRI) contrast agent candidate.[11]

MRI is a powerful noninvasive diagnostic technique for visualizing the fine structure of the human body with high spatial resolution. To obtain accurate diagnosis, contrast agents are usually used to increase the MRI image quality by exerting an influence on the longitudinal ($T_1$), transverse ($T_2$) relaxation time and $T_2$-star ($T_2^*$), a relaxation parameter arising principally from local magnetic field in homogeneities that are increased with iron deposition.[12] Due to the unique superparamagnetic property, $Fe_3O_4$ NPs have been widely used as ultrasensitive negative contrast agents for early detection of tumors due to their strong $T_2$ and $T_2^*$ shortening effect.[13]

Unfortunately, the exposed $Fe_3O_4$ NPs have poor solubility and biocompatibility. Therefore, the surface of $Fe_3O_4$ NPs needs to be suitably engineered to acquire improved biocompatibility.[14, 15] The most popular methods are introducing hydrophilic groups and modification of bioinorganic shell on the surface of $Fe_3O_4$ NPs, which can effectively optimize the properties of $Fe_3O_4$ NPs to satisfy the requirements of applications, such as MRI contrast agents, [16] magnetic targeting drug, biomolecule separation, and hyperthermal cancer therapy.[17] For example, $Fe_3O_4$-based nanocomposites, such as $Fe_3O_4$@Polydopamine,[18] $WS_2$@$Fe_3O_4$,[19] $Fe_3O_4$@Au[20] and C-$Fe_3O_4$ quantum dots (QDs)[21], have been studied as MRI contrast agents and drug carriers in preclinical and clinical setting.[22]

Among the modification materials of $Fe_3O_4$ NPs, polysaccharides have the advantages of stability, water solubility and few side effects on the organism. Meanwhile, some natural polysaccharides can promote the formation of hematopoietic cell.[23] Astragalus membranaceus, a traditional Chinese herb medicine containing many active components, such as polysaccharides, flavonoids, saponins, amino acids and trace elements, has been used in China for medicine for more than 2,000 years to improve human immunity and treat cardiovascular disorders.

Astragalus polysaccharide (APS) is one of the main active components of Astragalus membranaceus extracted from Astragalus roots, which is described in the 2005 version of the "Chinese Pharmacopoeia".[24] APS is comprised of uniform polysaccharide fraction, including rhamnose, arabinose and glucose.[25] APS is stable and biocompatible *in vivo*, which makes it an ideal material for biomedical applications. Many reports on the biological activities of APS indicate that it possesses potent anti-inflammatory, organ protective, anti-tumor activities and growth of red blood cell stimulative.[26] Therefore, APS is a promising material for modifying $Fe_3O_4$ NPs.

To develop a potential IDA therapeutic agent combining iron supplements and contents of promoting the hematopoietic cell formation, we report a novel material of $Fe_3O_4$@APS Core-Shell NPs fabricated by a two steps approach involving hydrothermal synthesis[27] and esterification. As shown in **Figure 1a**, the water soluble $Fe_3O_4$ NP cores were obtained in sodium citrate solution, in which sodium citrate replaces the oleic acid molecular on the surface of the $Fe_3O_4$ NPs. Then, The $Fe_3O_4$ NPs were bonded with the APS backbone, and the sodium citrate reacted with the hydroxyl groups of the APS chains to form the $Fe_3O_4$@APS NPs. Finally, the $Fe_3O_4$@APS NPs were used in mice intragastric administration for IDA treatment and MRI testing followed by pathological tests, after the structural stability characterization and cytotoxicity assessment, as depicted in **Scheme 1**. As expected, the $Fe_3O_4$@APS NPs show unique therapeutic effect for IDA treatment, which is much better than that of water soluble $Fe_3O_4$ NPs. This is the result of the combined contribution of the hematopoietic cell generation stimulated by APS and the simultaneous iron element supplement from $Fe_3O_4$. Besides this, the $Fe_3O_4$@APS NPs are also potential MRI contrast agents *in vivo*, which are safe and good for health. The study opens a new way to design IDA medicine with USPIO NPs and natural polysaccharides, which can serve as a contrast agent at the same time.

The crystal structural of the $Fe_3O_4$ NPs and $Fe_3O_4$@APS NPs were investigated by X-ray diffraction (XRD) (Figure S2, Supporting Information). The XRD patterns of both NP samples show sharp diffraction peaks of $Fe_3O_4$ (JCPDS75-0033)[28] with Fd-3m space group, indicating good crystallinity. Figure 1b shows the transmission electron microscope (TEM) image of the oil-soluble $Fe_3O_4$ NPs, which are spherical with a uniform diameter of ~10 nm. High-resolution TEM images of the $Fe_3O_4$ NPs exhibit (220) and (331) crystal facets as indexed in Figure 1c. The hydrodynamic diameter of the water soluble $Fe_3O_4$ NPs was measured to be 11 nm by dynamic light scattering (DLS) (Figure 1e), which is slightly larger than that measured by TEM (Figure 1d).

For the $Fe_3O_4$@APS NPs, the diameter does not show obvious increase in TEM image (Figure 1f), revealing clear cores of $Fe_3O_4$ NPs but insignificant evidence of the APS shells. This is because APS molecules are composed by C, H, O and N elements, which are invisible in TEM. However, the DLS results show a mean size of 29.5 nm for the $Fe_3O_4$@APS NPs (Figure 1g), which demonstrates the existence of the APS shells coated on the surface of $Fe_3O_4$ NPs. The dispersion stability of the oleic acid-coated $Fe_3O_4$ NPs shows good water stability and biocompatibility after the ligand exchange reaction on the surface from oleic acid to sodium citrate and APS. The zeta potential of the water soluble $Fe_3O_4$ NPs and $Fe_3O_4$@APS NPs are -36.8 mV and -28.8 mV, respectively, which suggests good water dispersity of the NPs (Figure S3, Supporting Information).[29]

Fourier transform infrared (FT-IR) absorption spectra were measured to confirm the presence of the APS around the $Fe_3O_4$ NPs (Figure 1h). Compared to the FT-IR spectrum of $Fe_3O_4$ around with oleic acid, the one of $Fe_3O_4$@APS NPs shows new strong absorption bands at 1383 cm$^{-1}$ and 1633 cm$^{-1}$, which can be attributed to the C-O-C and C=O stretch of APS and sodium citrate vibration of polysaccharide, respectively. It is evident that the oleic acid on the surface of $Fe_3O_4$ NPs was efficiently replaced by APS and sodium citrate. Moreover, the $Fe_3O_4$@APS NPs show a strong absorption band at 586 cm$^{-1}$, attributed to Fe-O bond, while the APS copolymer has the stretching bands at 2921 cm$^{-1}$ and 3424 cm$^{-1}$, corresponding to hydroxyl groups, which are a further evidence of the ligand exchange on the surface of $Fe_3O_4$ NPs.

To study the magnetic properties of $Fe_3O_4$@APS NPs, the magnetization curves have been measured, as shown in Figure 1i. All the curves exhibit no hysteresis, suggesting that the $Fe_3O_4$ NPs are superparamagnetic, with or without APS coating. The $Fe_3O_4$ NPs have a higher saturation magnetization ($M_s$) (about 62.2 emu/g) than that of the $Fe_3O_4$@APS NPs (about 36.7 emu/g), which further demonstrates the existence of the APS shells on the surface of $Fe_3O_4$ NPs. From the change of $M_s$, the mass percentage of the $Fe_3O_4$ NPs can be estimated to be around 60% in the

Fe$_3$O$_4$@APS NPs. With a magnet placed outside the cuvette, the Fe$_3$O$_4$@APS NPs in the water solution can rapidly accumulate near the magnet (See the inset of Figure 1i). When the magnet is removed, the Fe$_3$O$_4$@APS NPs disperse in water again, even after 6 months, suggesting the Fe$_3$O$_4$@APS NPs are superparamagnetic with good stability. To further investigate the magnetic behavior of the Fe$_3$O$_4$@APS NPs, the temperature dependence of their magnetization was investigated by zero-field cooling (ZFC) and field cooling (FC) respectively (Figure S4, Supporting Information). The M$_{ZFC}$-T curves show broad peaks with a clear maximum ($T_{max}$) at ~226 K for the Fe$_3$O$_4$ NPs and ~167 K for the Fe$_3$O$_4$@APS NPs, above which temperature the particles are in superparamagnetic regime. Thus, both kinds of the NPs are superparamagnetic at 300 K.[14]

Since the Fe$_3$O$_4$@APS NPs should be fed to the Institute of Cancer Research (ICR) mice and Wistar rats, it is possible that the Fe$_3$O$_4$@APS NPs might be degraded during digestion in their stomachs. Thus, we have checked the stability of the Fe$_3$O$_4$@APS NPs in imitated gastric acid by stirring in HCl solution with pH = 1.5 for 5 h. The magnetization curve of the Fe$_3$O$_4$@APS NPs collected after the imitating digestion still shows good superparamagnetic property (as shown in Figure 1i). The $M_s$ of the Fe$_3$O$_4$@APS NPs increases to 45.6 emu/g, which is larger than that of the as-prepared Fe$_3$O$_4$@APS NPs. This suggests that a certain amount of APS molecules could desorb from the NPs surfaces in the stomach due to the acid environment. However, comparing with the $M_s$ of the Fe$_3$O$_4$ NPs, we confirmed that there is still sufficient amount of APS molecules attached to the Fe$_3$O$_4$ NPs cores. Moreover, the FT-IR spectrum of the $Fe_3O_4$@APS NPs (Figure 1h) shows no significant change relative to that of the as-prepared $Fe_3O_4$@APS NPs, consistent with the results of the magnetic studies. The iron concentration in the $Fe_3O_4$@APS NPs was also quantitatively determined by inductively coupled plasma optical emission spectroscopy (ICP-OES). The weight percentage of Fe increases from 60% to 75% after stirring in HCl solution, which is consistent with the change of the magnetization curves. Accordingly, only a small amount of APS molecules were degraded by HCl solution. Therefore, the $Fe_3O_4$@APS NPs are stable enough and can be absorbed into blood as whole NPs, instead of the separate $Fe_3O_4$ NPs and APS.

The cytotoxicity evaluation is necessary for a new biomaterial before *in vivo* applications. We assessed the biocompatibility of the $Fe_3O_4$@APS NPs with mammalian cells by a 3-(4,5-dimethylthiazol-2-yl)-2,5-diphenyltetrazolium bromide (MTT) assay using a HeLa cell line. As shown in **Figure 2a**, the cell viabilities are higher than 75% after 24 h, 48 h and 72 h exposure to the iron concentration of 0-50 $\mu$g/mL. There is no decrease in cell viability when the exposure time was prolonged to 72 h. Therefore, the $Fe_3O_4$@APS NPs have a high biocompatibility and low cytotoxicity.

The MRI performance of the $Fe_3O_4$@APS NPs as contrast agents has been studied both *in vitro* and *in vivo*. For comparison, the $Fe_3O_4$ NPs have also been investigated. Figure 2b shows the $T_2^*$ MRI images of the $Fe_3O_4$ NPs and $Fe_3O_4$@APS NPs suspensions at different Fe concentrations. The low signal intensities in the $T_2^*$ MRI images increase with Fe concentration for both kinds of nanoparticles. Moreover, the intensities are similar when the Fe concentration in the $Fe_3O_4$ NPs and $Fe_3O_4$@APS NPs suspensions are approximately equal. As a negative MRI contrast agent, the superparamagnetic $Fe_3O_4$ core can shorten the $T_2^*$ values of water, resulting in a hyperintense signal on $T_2^*$ imaging. In addition, the $T_2^*$ relaxation rate ($r_2^* = 1/T_2^*$) of the $Fe_3O_4$ NPs and $Fe_3O_4$@APS NPs was calculated to be 483 $mM^{-1}s^{-1}$ (Figure 2c) and 400 $mM^{-1}s^{-1}$ (Figure 2d), respectively. The observed decrease in the $r_2^*$ value can be attributed to the APS on the surface of $Fe_3O_4$ NPs, which weakens the magnetic properties of the NPs. However, the $Fe_3O_4$@APS NPs remain a good contrast agent for MRI applications.

The preliminary *in vitro* MRI signal contrast enhancement of the $Fe_3O_4$@APS NPs led us to evaluate its possibility to serve as a negative MRI contrast agent for MRI *in vivo* imaging. $Fe_3O_4$@APS NPs were gastrically infused into the stomach of live healthy male ICR mice at a dosage of 10 mg/kg. Figure 2e shows the $T_1$ and $T_2$ weighted MRI images of the reference group (normal mice) and the mice after gastric infusion for 15 min. Comparing to the reference group, the $T_1$ and $T_2$ images for the mice after the gastric infusion show a positive and a negative contrast enhancement in the stomach and the bowels, respectively. Therefore, the $Fe_3O_4$@APS NPs have a significant signal enhancement effect as a MRI contract agent. The MRI images of the main organs have no specific changes even after being perfused for 4 h, 8 h and 16 h (Figure S5, Supporting Information), indicating the safety during $Fe_3O_4$@APS NPs metabolism.

*In vivo* toxicity and possible side effects of nanomedicines have to be carefully studied before practical applications in clinic. In order to further demonstrate the safety of the $Fe_3O_4$@APS NPs, the stomachs of healthy male ICR mice were perfused with the $Fe_3O_4$@APS NPs at a dosage of 1.0 mg/kg/d and 2.0 mg/kg/d, respectively. On the 15th and 30th day, the mice were tested by MRI to examine the effect of the $Fe_3O_4$@APS NPs on the organs such as liver and kidney. **Figure 3a** and Figure S6a (Supporting Information) show the $T_2$-weighted images of the livers and kidneys of the mouse treated by 1.0 mg/kg/d and 2.0 mg/kg/d of $Fe_3O_4$@APS NPs on the 30th and 15th day, together with that of the normal mouse as contrast group. Comparing with the contrast group, there is no obvious damage in the groups treated by $Fe_3O_4$@APS NPs. The $T_2^*$ values shown in Figure 3b and Figure S6b (Supporting Information) also reveal no particular change in $T_2^*$ values of the kidney and the liver in the range of the organ regions with different dosage and time. These results demonstrate that no pathological change happened in the organs of the ICR mice after a long time supplementation of the $Fe_3O_4$@APS NPs.

The histopathological studies were carried out on the main organs of the ICR mice after the *in vivo* MRI testing. The histopathological images of the livers, kidneys and brains are presented in Figure 3c and Figure S7 (Supporting Information), which correspond to the ICR mice after the intragastric administration of $Fe_3O_4$@APS NPs for 30 and 15 days, respectively. As expected, no particles were observed in the organs, and no sign of organ damage or inflammation was observed, demonstrating the minimal side effects of the $Fe_3O_4$@APS NPs *in vivo* after 30 days intragastric administration.

To study the effect of the $Fe_3O_4$@APS NPs on IDA, low-iron diet Wistar rats were separated into different groups and fed with the designed ferralia dosage. The rats were weighted every week. As shown in Figure 3d, the mean body weights of all groups increase almost linearly with time. As expected, the model group shows the smallest slope of the weight growth due to serious IDA. The group treated with the $Fe_3O_4$ NPs has a weight growth rate higher than the model group but lower than the other treated and normal groups without IDA. However, the groups treated with the $Fe_3O_4$@APS NPs have weight growth rates comparable to that of the normal group, and a higher $Fe_3O_4$@APS NPs dosage results in a larger weight growth rate. Therefore, the $Fe_3O_4$@APS NPs treatment can improve the health of the low-iron diet rats.

The Hemoglobin (HGB), Hemoglobin (RBC) and Hematocrit (HCT) are important parameters for the diagnosis and therapy of IDA. Therefore, the blood analyses were carried out for the rats studied, and the results are presented in Figure 3e-3g. Usually, the rats are considered as anemic when the HGB level is reduced to 75% of the original level. As shown in Figure 3e, the HGB values of the low-iron diet groups are only half of that of the normal group before ferralia treatment, which indicates a serious IDA of the low-iron diet rats. After the ferralia treatment, all the treated groups have significant increase of HGB values, while only the group supplied with the $Fe_3O_4$@APS NPs at a dosage of 2.0 mg/kg/d has a HGB value (136 g/L) comparable to the normal group (143.1 g/L). The group treated with the 1.0 mg/kg/d $Fe_3O_4$@APS NPs (117.1 g/L) has a curative effect similar to the group of the 2.0 mg/kg/d $Fe_3O_4$

NPs (117.2 g/L), demonstrating that the $Fe_3O_4$@APS NPs have more positive effect on HGB than the $Fe_3O_4$ NPs. The other two important parameters, RBC and HCT, have the trends similar to that of HGB. All blood analyses suggest that the health condition of the group supplied by the 2.0 mg/kg/d $Fe_3O_4$@APS NPs recovered to the level of the normal group. Moreover, the superoxide dismutase (SOD), glutathione (GSH), lactic dehydrogenase (LDH), hepcidin (HEP), malondialdehyde (MDA) and serum iron (SI) data were also measured after 30 days of the treatment. All results (Table S1, Supporting Information) demonstrate the high therapeutic effect of the $Fe_3O_4$@APS NPs on IDA *in vivo*. Considering the structure of the $Fe_3O_4$@APS NPs, their high therapeutic effect is attributed to both iron element supplement from $Fe_3O_4$ and the APS stimulated hematopoietic cell generation. As a result, the $Fe_3O_4$@APS NPs are highly promising as an IDA drug with the additional benefit of serving as a MRI contrast agent.

In conclusion, we have designed and experimentally demonstrated the $Fe_3O_4$@APS NPs for the targeting IDA treatment and the MRI contract agent. The stability of the $Fe_3O_4$@APS NPs was determined by the imitating digestion in imitated gastric acid. Both *in vitro* and *in vivo* toxicity studies were carried out using the cell experiments and intragastric administration of animal model. The blood analyses show that the $Fe_3O_4$@APS NPs have potent therapeutic effect on IDA, evidenced by the HGB, RBC and HCT values of the 2.0 mg/kg/d $Fe_3O_4$@APS NPs treated IDA rats. Moreover, strong MRI contrast enhancement has been observed for the $Fe_3O_4$@APS NPs. Therefore, the $Fe_3O_4$@APS NPs are promising candidates for IDA drugs with the additional functionality as MRI contract agent.

## Supporting Information

Supporting Information is available from the Wiley Online Library or from the author.

## Acknowledgements

K. Wang and L. N. Li contributed equally to this work. This work was partially supported by the National Basic Research Program of China (Grant No. 2015CB921502), the National Science Foundation of China (Grant Nos. 51671019, 51471029, 51731003, 61471036) and Beijing science and technology Nova cross program (Z171100001117136).

## Conflict of Interest

The authors declare no conflict of interest.

## Keywords

$Fe_3O_4$ nanoparticles, astragalus polysaccharides, iron deficiency anemia, magnetic resonance imaging.

**Figure Captions**

**Scheme 1.** Schematic illustration of the design and application of $Fe_3O_4$@APS NPs for IDA therapy and MRI.

**Figure 1.** a) Schematic drawing the synthesis of the $Fe_3O_4$@APS NPs. b) TEM and c) high-resolution TEM images of the oil soluble $Fe_3O_4$ NPs. d) TEM image of water soluble $Fe_3O_4$ NPs and e) their hydrodynamic profiles. f) TEM image of $Fe_3O_4$@APS NPs and g) their hydrodynamic profiles. h) FT-IR spectra and i) magnetization curves of the $Fe_3O_4$ NPs, $Fe_3O_4$@APS NPS and $Fe_3O_4$@APS NPs after stirring in HCl solution. Inset of i) is the image of the $Fe_3O_4$@APS NPs in water dispersion with (right) and without (left) external field.

**Figure 2.** a) Cell viability of Hela cells after treatment with the $Fe_3O_4$@APS NPs at different iron concentrations after 24h, 48h and 72h. b) $T_2^*$ images of the $Fe_3O_4$ NPs and $Fe_3O_4$@APS NPs water solution with different Fe concentrations. Corresponding $T_2^*$ relaxation rate of the $Fe_3O_4$ NPS c) and the $Fe_3O_4$@APS NPs d) with respect to iron concentration. e) $T_1$-weighted and $T_2$-weighted MRI images of contrast group (normal mouse) and the $Fe_3O_4$@APS NPs gastric infused Male ICR mouse after 15 min.

**Figure 3.** a) *In vivo* $T_2$-weighted images of the male ICR mice liver and kidney for contract group and the group treated with the $Fe_3O_4$@APS NPs at different Fe concentrations after 30 days. b) Signal intensity of $T_2^*$ values of liver and kidney. c) H&E stained organ slices of the mice after 30 days $Fe_3O_4$@APS NPs treatment. Scale bars for all images are 100 $\mu$m. d) Body Weight increase with time, e) HGB concentration, f) RBC concentration and g) HCT values of male Wistar rats (8 per group) before and after 30 days treatment with the $Fe_3O_4$@APS NPs intragastric administration.

Scheme 1. (Wang K. et al.)

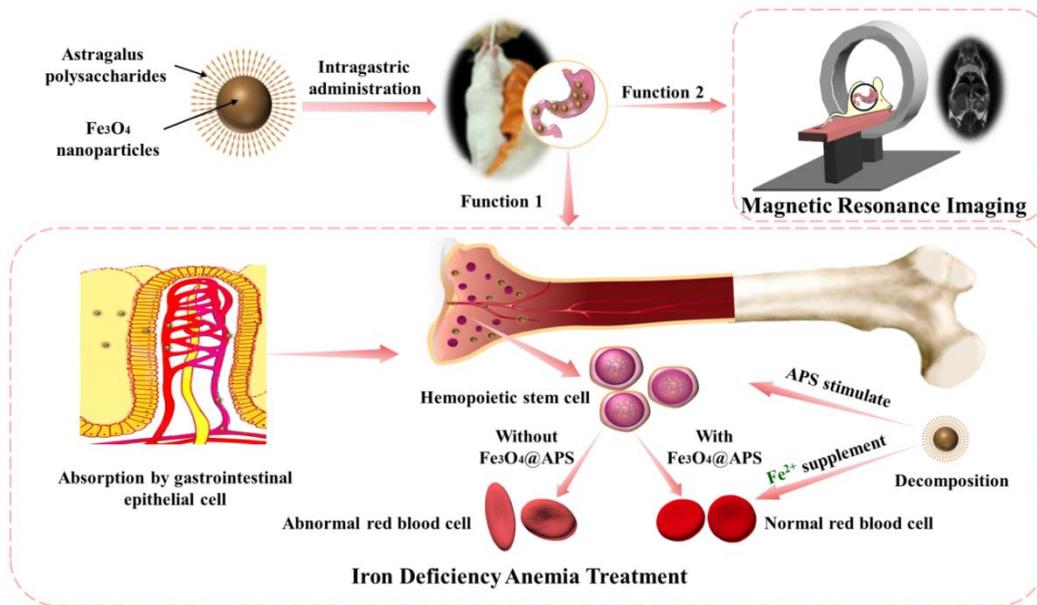

Figure 1. (Wang K. et al.)

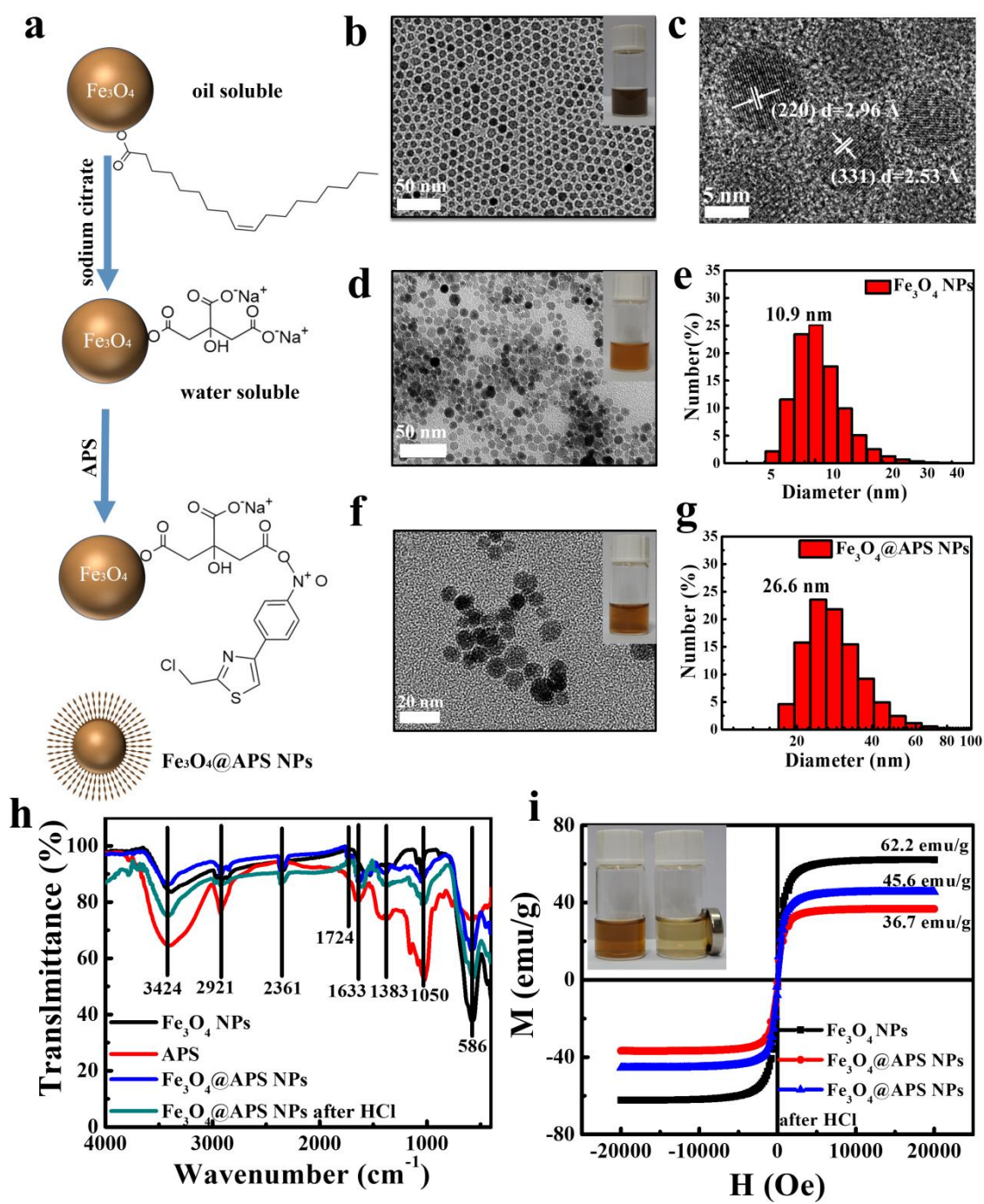

Figure 2. (Wang K. et al.)

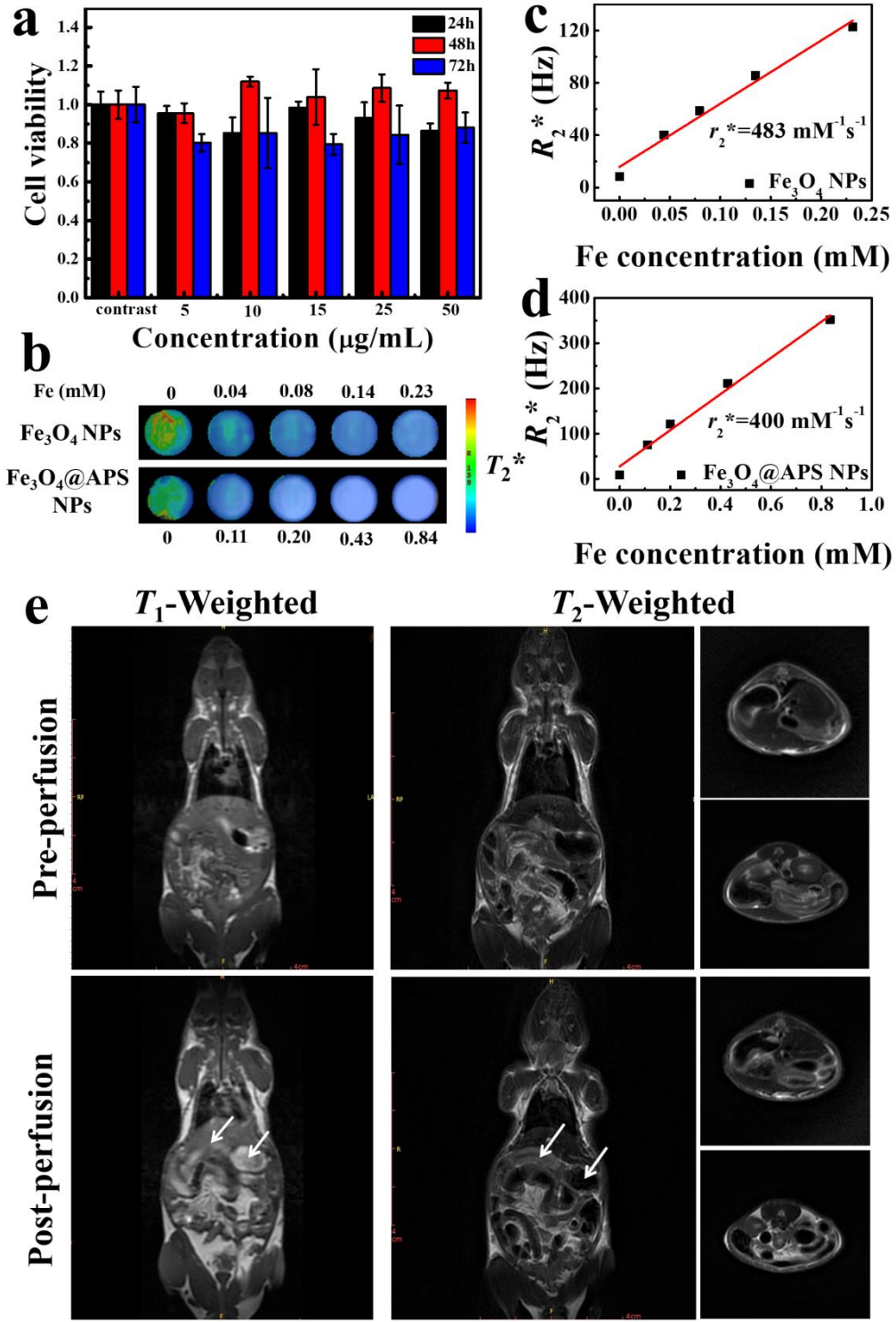

Figure 3. (Wang K. et al.)

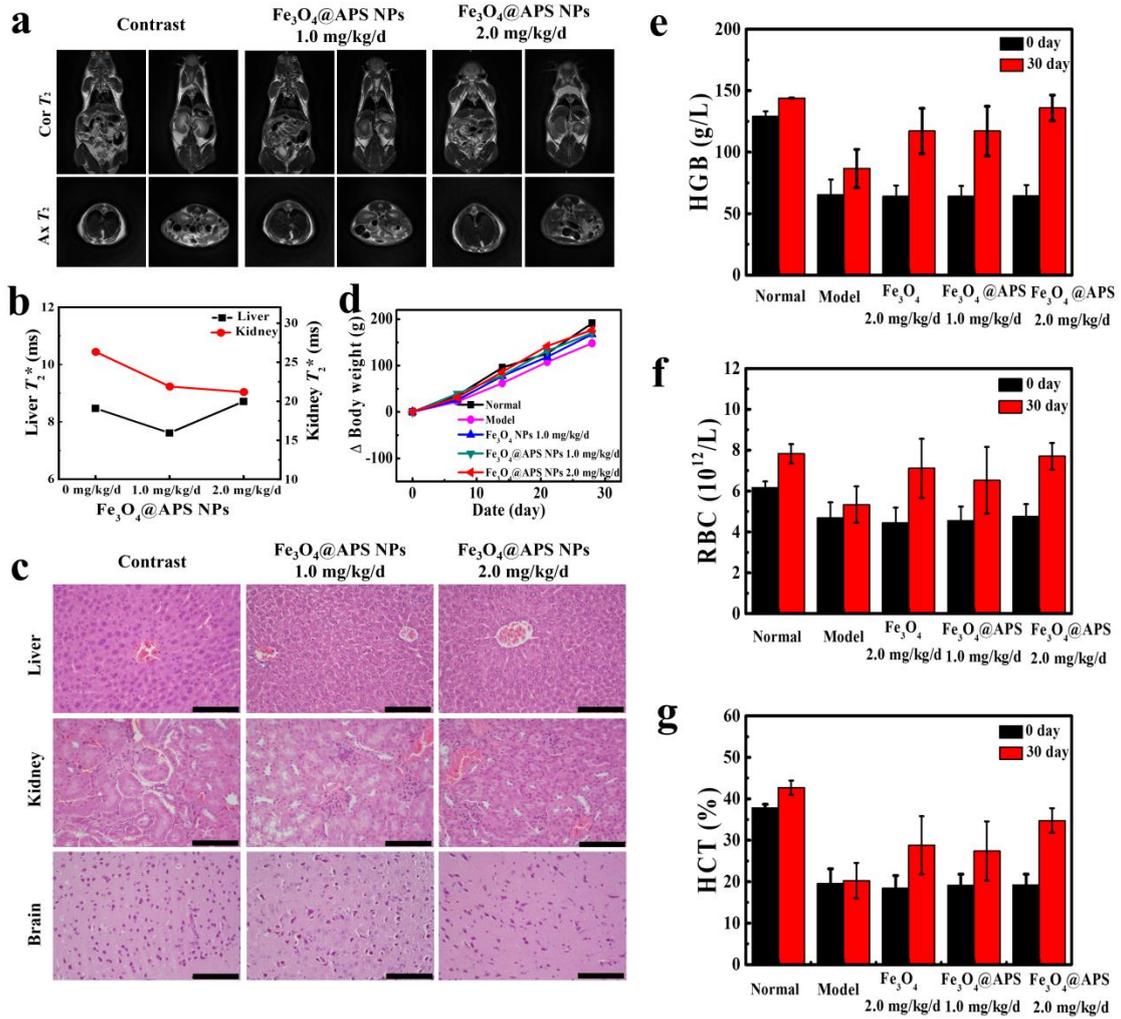